\documentstyle[preprint,aps,prd,tighten,amssymb,eqsecnum,epsf]{revtex}
\begin{document}
\draft
\title{Exact Solutions for the Intrinsic Geometry of Black Hole Coalescence}

\author{Luis Lehner${}^{1}$\thanks{Present Address:
                                   Center for Relativity,
                                   The University of Texas at Austin,
                                   Austin TX 78712.},
	Nigel T. Bishop${}^{2}$,
	Roberto G\'omez${}^{1}$, \\
	Bela Szilagyi${}^{1}$ and
	Jeffrey Winicour${}^{1}$
	}
\address{
${}^{1}$Department of Physics and Astronomy,\\
University of Pittsburgh, Pittsburgh, PA 15260 \\
${}^{2}$Department of Mathematics, Applied Mathematics and Astronomy,\\
University of South Africa, P.O. Box 392, Pretoria 0003, South Africa}
\maketitle

\begin{abstract}

We describe the null geometry of a multiple black hole event horizon in
terms of a conformal rescaling of a flat space null hypersurface. For
the prolate spheroidal case, we show that the method reproduces the
pair-of-pants shaped horizon found in the numerical simulation of the
head-on-collision of black holes. For the oblate case, it reproduces
the initially toroidal event horizon found in the numerical simulation
of collapse  of a rotating cluster. The analytic nature of the approach
makes further conclusions possible, such as a bearing on the hoop
conjecture. From a time reversed point of view, the approach yields a
description of the past event horizon of a fissioning white hole, which
can be used as null data for the characteristic evolution of the
exterior space-time.

\end{abstract}


\section{Introduction}
\label{sec:int}

Numerical simulations of axisymmetric space-times have enabled
construction of the event horizon traced out by the evolution of
dynamical black holes~\cite{shteuk,annin1,torus,ccode,annin2,libson}.
Besides confirming behavior dictated by the general laws of black hole
dynamics~\cite{hawkell,wald}, the simulations supply further insight
which was not anticipated from analytic theory. In the case of a
head-on collision, they supply the details of how the holes form and
merge~\cite{sci}. In the case of rotating collapse, they reveal how an
initially toroidal structure is compatible with topological
censorship~\cite{toroid}. In this paper, we present a 3-dimensional
analytic description of the event horizon for a multiple black hole
space-time which, in the axisymmetric case, reproduces the qualitative
features of the above simulations.

As a stand-alone item, a black hole horizon is a null hypersurface
whose cross-sectional surface area monotonically increases and
approaches a finite limit in the future. The {\em number} of black
holes contained at a given time is not conventionally defined in terms
of such a stand-alone picture but rather in terms of the number of
disjoint sections given by the intersection of the horizon with a
Cauchy hypersurface~\cite{hawkell,wald}. In the approach we present
here, in the case of the head-on-collision the notion of ``two
holes'' arises intrinsically from a preferred slicing of the horizon
based upon an affine parameter along its generating null rays.

For this purpose, we consider the geometry of a null hypersurface
${\cal N}$ whose surface area has a finite asymptotic limit. The
intrinsic and extrinsic geometrical properties of a null hypersurface
cannot be described in terms of the same ``3+1'' formalism used for a
space-like hypersurface. In particular, the degenerate 3-metric of a
null hypersurface does not define an intrinsic covariant derivative,
in contrast to the case of a space-like hypersurface.
Dautcourt~\cite{daut} has presented an alternative formalism for
inducing an intrinsic geometry on a null hypersurface from the
embedding geometry.  Geroch~\cite{cinn} used a similar approach to
describe null infinity ${\cal I}$ as a 3-dimensional null
hypersurface detached from the physical space-time $M$. In Sec.
\ref{sec:proj}, we take an analogous approach to treat the null
hypersurface traced out by a black hole as a stand-alone geometric
object.

Numerical investigations~\cite{shteuk,annin1,torus,ccode,annin2,libson} of
event horizons have used the Cauchy initial value problem to evolve a black
hole space-time throughout a domain of sufficient extent (both in space and
time) so as to include an apparent horizon which has become approximately (but
not exactly) stationary. This implies that the marginally trapped surfaces
defining the apparent horizon have almost stopped growing in surface area. The
event horizon is then located by examining the null geodesics which pass in the
vicinity of the last quasi-stationary marginally trapped surface obtained in
the evolution. The approach adopted in this paper is rather different. We have
developed a method that constructs a null hypersurface whose surface area has
finite asymptotic limit and whose intrinsic geometry satisfies all other
requirements for a non singular event horizon. Thus, from all intrinsic
criteria, this null hypersurface represents an event horizon ${\cal H}$. The
results of this paper depend only upon the intrinsic geometry of ${\cal H}$ and
not upon the properties of the embedding space-time (which will be presented in
subsequent papers). Even so, in order to show that the results of this paper
are physically meaningful, it is necessary to discuss whether there exists a
space-time in which ${\cal H}$ can be embedded.

\begin{figure}
\centerline{\epsfxsize=6in\epsfbox{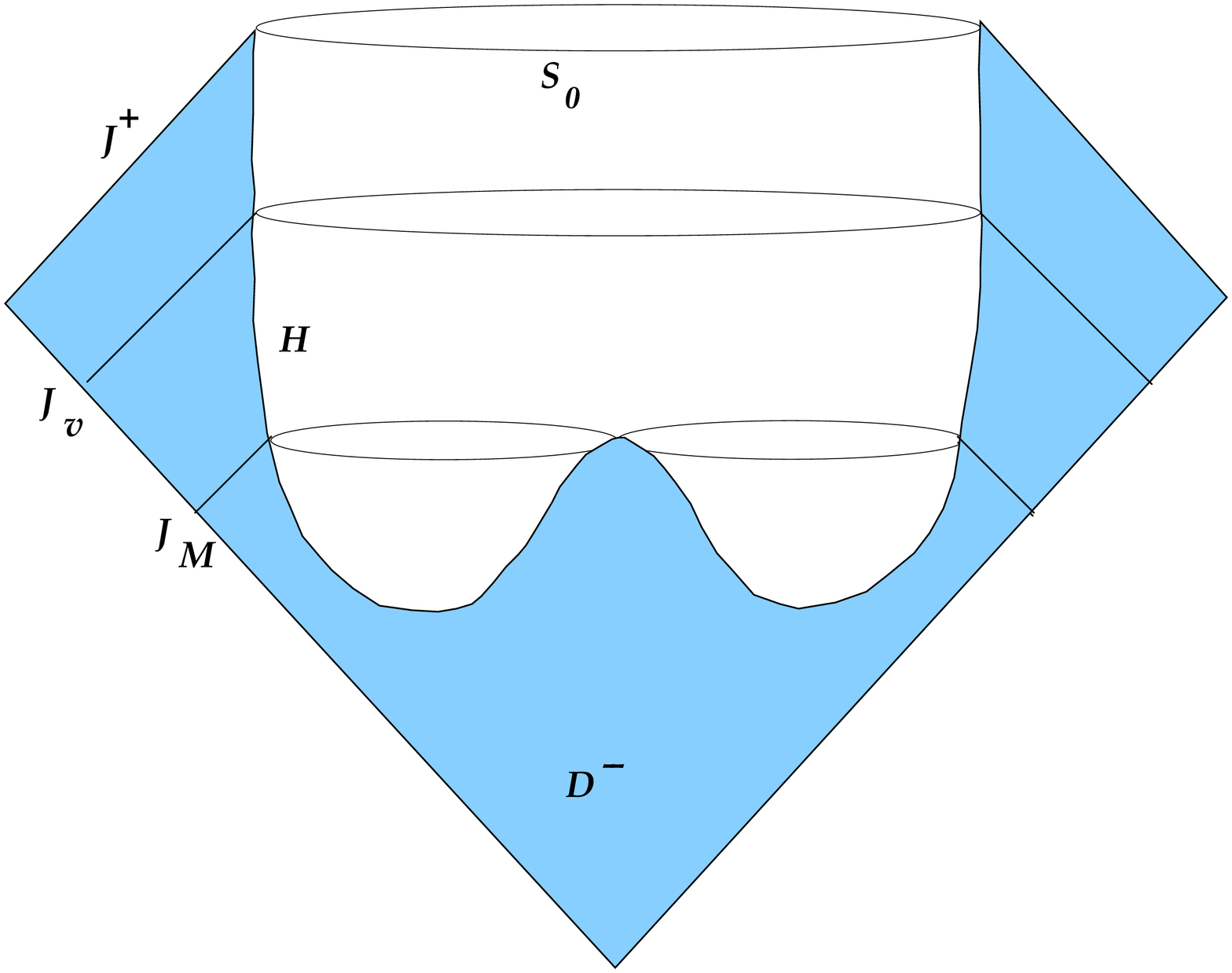}}
\caption{A portion of the space-time prior to the coalescence of two black
holes. The parameter $v$ labels the advanced time on a family of incoming
null hypersurfaces. $D^-$ is the domain  of dependence of characteristic data
given on the event horizon ${\cal H}$ and on $J^+$.}
\label{fig:civp}
\end{figure}

There is a formal construction of a vacuum space-time based upon the
characteristic initial value problem posed on an intersecting pair of null
hypersurfaces~\cite{sachsdn,haywsn,haywdn}. Here we consider two null
hypersurfaces intersecting in a topologically spherical surface $S_0$.
Intrinsic geometrical data must be given on the two null hypersurfaces, as
well as a quantity, called the twist \cite{haywsn,haywdn}, which is analogous
to an extrinsic curvature in the Cauchy problem. We can apply this formalism
to the case where $S_0$ is a cross-section of the event horizon at a late
quasi-stationary time. Then, referring to Fig.~\ref{fig:civp}, ${\cal H}$
intersects the ingoing null hypersurface $J^+$ at $S_0$. Characteristic data
given on the portion of ${\cal H}$ and $J^+$ to the past of $S_0$, leads to a
formal solution of the vacuum equations in the domain of dependence $D^-$ to
the past of $S_0$. To the extent that $S_0$ lies in the late time stationary
region of the horizon, $J^+$ approximates future null infinity. So, for a
horizon describing a binary black hole, the physically appropriate
characteristic data on $J^+$ would describe the outgoing radiation emitted
during the merger, although any data satisfying the constraints would
formally lead to a consistent vacuum space-time. (As in the Cauchy problem,
Einstein's equations imply constraints  on the characteristic initial data
but these reduce to propagation equations along the null geodesics so that it
is simple to isolate the unconstrained free data.)

There are theorems establishing the existence of solutions to this double null
initial value problem~\cite{hagenseifert77,helmut81a,helmut81b}. Although
global issues remain unresolved, these theorems guarantee existence in some
neighborhood of the initial hypersurfaces. The double null version of the
characteristic initial value problem is equivalent to the world tube - null
cone problem~\cite{tamb} in the case where the world tube is null. There exists
a stable, accurate and efficient characteristic evolution
code~\cite{highp,wobb} (the PITT code) which evolves this initial value problem
and which could be applied to construct a numerical solution to the  situation
described in Fig.~\ref{fig:civp}.  This provides the local existence of a
space-time satisfying Einstein's equations in the finite difference
approximation. Ideally, we would like to construct a numerical solution
throughout the domain of dependence $D^-$ but that is more problematical. The
evolution algorithm requires the foliation of $D^-$ by a one parameter family
of null hypersurfaces $J_v$. However, referring to Fig.~\ref{fig:civp}, such a
foliation becomes singular at $J_M$ in the portion of the horizon corresponding
to black hole merger. Thus one could only obtain the post-merger space-time by
this procedure. This problem is due to a coordinate singularity, not a
physical singularity, and from a mathematical point-of-view the space-time is
extendable to earlier times; but just how much earlier cannot be answered by a
characteristic evolution.  Thus the analytic event horizon for a black hole
collision, which we present in this paper, can be used as characteristic
initial data to construct a vacuum space-time (either analytically or
numerically) covering some domain preceding the merger.

A major motivation for the present study stems from a potential indirect
application of the results to a calculation of the gravitational waveform
radiated by coalescing black holes using the PITT code. Waveforms from highly
nonlinear, highly distorted single black holes have already been obtained
with this code~\cite{highp,wobb}. In these simulations~\cite{highp}, the two
null hypersurfaces were chosen to be (i) a portion of the past horizon (white
hole) of a Schwarzschild space-time and (ii) an outgoing null hypersurface
emanating from a slice of the past horizon to ${\cal I}^+$, which (in order
to introduce distortion) contains incoming radiation. The code has equal
capability of carrying out such simulations with the static Schwarzschild
white hole replaced by a dynamic past horizon corresponding to a white hole
which is initially stationary and later fissions into a pair of white holes.
In a time-reversed scenario, the outgoing waveform from this white hole
explosion corresponds to the {\em incoming} radiation incident from past null
infinity on the merger of a pair of black holes. While this is not the
correct physical prescription of initial conditions for black hole
coalescence, if the system were linear this incoming radiation could be used
to draw inferences about the {\it outgoing} radiation. Such linearity is of
course not expected. Yet any means of obtaining a handle on the merger
waveform is of current importance. In addition, a solution of this problem
would unambiguously yield the outgoing radiation from a white
hole explosion, a system of at least academic interest.

The time reversal of the intrinsic geometry of the event horizon of a
black hole, results in the event horizon of a white hole, and vice-versa.
The construction developed in this paper is naturally
expressed in terms of a white hole that bifurcates in the future, and thus
our method is presented for the case of a white hole. The reinterpretation
of our results, in terms of a black hole horizon, is straightforward.
Thus we pose our investigation in terms of a white hole horizon ${\cal
H}$, which constitutes a portion of a null hypersurface ${\cal N}$,
with the property that its surface area decreases into the future and
has a a finite asymptotic limit in the past. The null rays of ${\cal
N}$ leave ${\cal H}$ at points where they meet other null rays of
${\cal N}$. Such endpoints can occur at a ``crossover'' point, where
initially non-neighboring rays intersect, or at a caustic, where
neighboring rays intersect. These properties follow from the
fundamental theorems of black hole physics.~\cite{hawkell,wald}

Note that a crossover point may also be a caustic, as in the case of a
spherically symmetric light cone and also in the case of the prolate
spheroidal light cone considered in Section \ref{sec:axi}. In addition,
a portion of ${\cal N}$ can continue smoothly across a non-caustic
crossover point, as in the example of the oblate spheroidal light cone
considered in Section \ref{sec:axi}. However, ${\cal H}$ must end at
such a crossover point.

In this paper, we do not calculate the gravitational radiation emitted by the
system but study only the internal dynamics of ${\cal H}$. In Sec.
\ref{sec:axi} we specialize to the axisymmetric case which facilitates an
analytic treatment of the endpoints of ${\cal H}$. We find that the
bifurcation of ${\cal H}$ in a white hole explosion has the same
``pair-of-pants'' structure (in a time reversed sense) observed in the
numerical simulation of a head-on collision of black holes. Further features
emerge, such as the ultimate fate of an ``eternal'' pants leg, the details
of toroidal black hole formation in the vacuum case
and a bearing on a strict version of the hoop conjecture~\cite{hoop}.

\section{3-dimensional affine null geometry}
\label{sec:proj}

It is useful to consider ${\cal N}$  as one of the null hypersurfaces
in the double null initial value problem for the vacuum Einstein
equations. As first shown by Sachs~\cite{sachsdn}, the evolution of the
double null problem requires as boundary data the intrinsic conformal
geometry on ${\cal N}$, i.e. a metric $\gamma_{ab}$  expressed in terms
of an affine parameter $u$, up to the conformal freedom $\gamma_{ab}
\rightarrow \Omega^2 \gamma_{ab}$. In addition, the intrinsic conformal
geometry must be specified on a null hypersurface meeting ${\cal N}$
transversely at some cross-section ${\cal S}_0$, as well as the
intrinsic 2-geometry and certain extrinsic curvature quantities of
${\cal S}_0$. Here we restrict our attention to the intrinsic
properties of ${\cal N}$ and ${\cal S}_0$. As emphasized by
Hayward~\cite{haywsn,haywdn}, it is important to consider specification
of an affine parameter $u$ on ${\cal N}$ as part of the data.

This intrinsic data obey the Sachs equations~\cite{sachseq} governing
the expansion or contraction of ${\cal N}$. As a consequence, the data
determine a unique metric $\gamma_{ab}$ from the conformal equivalence
class on ${\cal N}$. Here $\gamma_{ab}$ satisfies the degeneracy
condition $\gamma_{ab} n^b =0$, where $n^b$ is tangent to the
generators of ${\cal N}$. We choose $n^a$ to have the affine
normalization $n^a \partial_a=\partial_u$.

Our aim is to use a
special choice of ${\cal N}$ as a stand-alone model of a white hole
horizon ${\cal H}$.  We require that ${\cal H}$ be complete in the past
and that its surface area have a finite asymptotic limit as $u
\rightarrow -\infty$.  ${\cal H}$ ends in the future at points where
its generators intersect, either at a caustic or crossover point.

The Sachs equations can be derived by projecting the relevant
components of the 4-dimensional Einstein equation into 3-dimensional
form. In doing so, we must deal with the degeneracy of $\gamma_{ab}$ in
representing the counterpart of 4-dimensional covariant derivatives as
operators on ${\cal N}$. The formalism adopted here makes explicit use
of the affine structure of ${\cal N}$.

We begin with the four dimensional description of ${\cal N}$ as a null
hypersurface embedded in a vacuum space-time with metric $g_{ab}$ and covariant
derivative $\nabla_a$.  On ${\cal N}$, the affine tangent to the generators
satisfies the geodesic equation $n^b\nabla_b n^a=0$ and the hypersurface
orthogonality condition  $n^{[a}\nabla^b n^{c]}=0$ We make no assumptions about
the behavior of $n^a$ off ${\cal N}$. We project 4-dimensional tensor fields
into ${\cal N}$ using the operator
\begin{equation}
       P_a^b = \delta_a^b + n_a l^b
\end{equation}
where $l^a$ is the unique outgoing null vector field on ${\cal N}$ which
is orthogonal to the affine cross-sections and satisfies $l^a n_a =-1$.
We can set $l_a = -\nabla_a u$, where $u$ is any smooth extension of the
affine parameter to a field in the neighborhood of ${\cal N}$.

The projected metric $\gamma_{ab}=
P_a^c P_b^d g_{cd}$ is the pullback of the 4-metric $g_{ab}$ to ${\cal
N}$.  When restricted to the 2-surfaces determined by the affine
foliation, the projected contravariant metric $\gamma^{ab}= P^a_c P^b_d
g^{cd}$ is the (unique) inverse of the pullback of $g_{ab}$.

We introduce the shorthand notation $\perp T_a^b$ for the projection
(to the tangent space of ${\cal N}$) of a tensor $T_a^b$. Thus, $\perp
n^a =n^a$ and $\perp l_a =l_a$. In addition, the following useful
formula hold on ${\cal N}$:
\begin{eqnarray}
	 0  & = & \perp n_a \\
	 0  & = & \perp l^a \\
	 0  & = & \perp {\cal L}_n l_a \\
	 0  & = & \perp {\cal L}_n n_a \\
	 0  & = & \perp \nabla_{[a} n_{b]} \\
	 \perp {\cal L}_n g_{ab}  & = & {\cal L}_n \gamma_{ab} \\
	 \perp {\cal L}_n^2 g_{ab}  & = & {\cal L}_n^2 \gamma_{ab}.
\end{eqnarray}
The last two of these equations can be verified using the commutation relation
\begin{equation}
	 0  = \perp [{\cal L}_n, \perp].
\end{equation}

Our purpose is to rewrite the projected curvature components
$\Phi_{ab}=\perp n^c n^d R_{cabd}=\perp n^c (\nabla_c \nabla_a-\nabla_a
\nabla_c) n_b$ in a form intrinsic to ${\cal N}$. By applying the above
formulae, we obtain
\begin{equation}
  \Phi_{ab} = \frac{1}{2}{\cal L}_n^2 \gamma_{ab} - \frac{1}{4}
  \gamma^{cd}({\cal L}_n \gamma_{ac}){\cal L}_n \gamma_{bd}.
\end{equation}
This further simplifies by setting $\gamma_{ab}=R^2 h_{ab}$ and
$\gamma^{ab}=R^{-2} h^{ab}$ where $h^{ab} {\cal L}_n h_{ab} =0$. (This
can be achieved by choosing $R^2$ as the determinant of the restriction
of $\gamma_{ab}$ to the surfaces of the affine foliation.) Then, in
terms of the shear tensor $\Sigma_{ab} ={\cal L}_n h_{ab}$,
\begin{equation}
  \Phi_{ab} = \frac{1}{2}{\cal L}_n (R^2 h_{ab})
	+h_{ab}R {\cal L}_n^2 R -  \frac{1}{4}R^2
   h^{cd}\Sigma_{ac}\Sigma_{bd}.  \label{eq:phiab}
\end{equation}

The Sachs equations follow immediately from Eq. (\ref{eq:phiab}).
Taking the trace of $\Phi_{ab}$ results in
\begin{equation}
n^c n^d R_{cd} =-\frac{2 {\cal L}_n^2 R}{R} - \frac{1}{2} \Sigma^2,
\end{equation}
where $R_{cd}$ is the Ricci tensor and where $\Sigma^2 =(1/2)
h^{ab}h^{cd}\Sigma_{ac}\Sigma_{bd}$. Then in the {\em vacuum case} it
follows that
\begin{equation}
   {\cal L}_n^2 R = -\frac{1}{4} R \Sigma^2.
\label{eq:focus}
\end{equation}
The trace free part of Eq. (\ref{eq:phiab}) yields
\begin{equation}
       \Psi_{ab} = \frac{1}{2}{\cal L}_n (R^2 \Sigma_{ab})
		 - \frac{1}{2}R^2 \Sigma^2 h_{ab}.
\label{eq:psiab}
\end{equation}
where $\Psi_{ab}$ are projected components of Weyl curvature.

We proceed further to decompose the shear in terms of its normalized
eigenvectors $p^a$ and $q^a$ satisfying $\Sigma_{ab}(p^b+iq^b)=\Sigma
h_{ab}(p^b-iq^b)$. Then $h_{ab}=p_ap_b+q_aq_b$ and $\Sigma_{ab}=\Sigma
(p_a p_b-q_aq_b)$, where $p_a+iq_a=h_{ab}(p^b+iq^b)$ and satisfies
$p^a{\cal L}_n p_a +q^a {\cal L}_n q_a =0$ and $p^a{\cal L}_n q_a +q^a
{\cal L}_n p_a =0$.

The goal is to solve Eq. (\ref{eq:focus}) in a way consistent with a
white hole horizon ${\cal H}$ on a portion of ${\cal N}$.  Thus the
Weyl curvature must be non-singular on ${\cal H}$. This requires that
the contraction $\Psi_{ab}$ with a unit vector (normalized with respect
to $\gamma_{ab}$) yield a non-singular scalar field on ${\cal H}$
(including the endpoints of ${\cal H}$); i.e. that
\begin{equation}
   \Psi = R^{-2}(p^a +iq^a)(p^b +iq^b)\Psi_{ab}
    =R^{-2}{\cal L}_n (R^2 \Sigma)
     +i\Sigma(q^a{\cal L}_n p_a -p^a{\cal L}_n q_a)
\label{eq:Psi}
\end{equation}
be non-singular.

\section{Conformally flat null geometry}
\label{sec:conflat}

The Weyl smooth solutions of Eq. (\ref{eq:focus}) have a large degree
of  freedom corresponding to the outgoing radiation crossing ${\cal
N}$. In order to restrict this freedom we consider solutions whose null
metric is conformal to that of a null hypersurface embedded in a flat
Minkowski space-time.

Consider then the flat space case, where we denote the corresponding
fields on ${\cal N}$ as ${\hat \gamma}_{ab}$, $\hat R$, ${\hat
h}_{ab}$, $\hat n^a$, $\hat u$, etc. For convenience, we write
$F'= {\cal L}_{\hat n} F$ for tensor fields $F$. Since $\hat \Psi_{ab}
=0$, Eq. (\ref{eq:psiab}) implies $(\hat p^a+i\hat q^a)(\hat p_a-i\hat
q_a)'=0$ and ${\hat R}^2 \hat \Sigma =\sigma$, where $\sigma '=0$. The
conditions on the eigenvectors may be summarized by $\hat p_a '=(\Sigma
/2)\hat p_a$ and $\hat q_a '=-(\Sigma /2)\hat q_a$.

The focusing equation (\ref{eq:focus}) now integrates to give
\begin{equation}
	 \hat R^2 =(A \hat u +B)^2-\frac{1}{4}\sigma^2,
\label{eq:1rhat}
\end{equation}
where $A'=B'=0$. We adjust the affine freedom in $\hat u$ so that
$\hat R\rightarrow \hat u$ as $\hat u \rightarrow \infty$
and so that the two caustics of $\hat R$ are placed symmetrically. Then
Eq. (\ref{eq:1rhat}) reduces to
\begin{equation}
 \hat R^2=(\hat u+\frac{1}{2}\sigma)(\hat u-\frac{1}{2}\sigma).
\label{eq:rhat}
\end{equation}
We choose conventions so that the $\sigma \ge 0$, so that the
caustic corresponding to the $q$ principle direction is reached
first, moving along a ray in the direction of increasing $\hat u$.

The dependence of the eigenvectors is determined to be
\begin{equation}
\hat p_a   = \bigg( \frac {\hat u -\sigma /2}
		   {\hat u +\sigma /2} \bigg)^{1/2} P_a
\end{equation}
\begin{equation}
\hat q_a   = \bigg( \frac {\hat u +\sigma /2}
		   {\hat u -\sigma /2} \bigg)^{1/2} Q_a
\end{equation}
where $(P_a+iQ_a)'=0$. The resulting $\hat u$-dependent family of
2-metrics, comprise the classic description of the 2-geometries
generated by the parallel map~\cite{parallel} of a surface ${\cal S}_0$
embedded in Euclidean space. The parallel map consists of translations
by the same distance $\Delta \hat u$ along each normal to the surface
(identical to Huyghen's construction for propagating a wavefront).
Equivalently, by considering ${\cal S}_0$ to be embedded at time $t=0$ in
a Minkowski space-time, the translation along each ingoing normal null
direction through the time $\Delta \hat u =t$, generates a
null-hypersurface foliated by constant time slices $\hat S_t$.

In applying this construction, we choose ${\cal S}_0$ to be convex so that
$\hat S_t$ traces out the flat space wavefronts of a null hypersurface
converging in the inward direction. From the point of view of the flat
embedding, $\sigma$ is the distance between the two caustics generically
encountered along each null ray.

Given such a flat space null hypersurface, with the convexity property
that its caustics are reached at finite $\hat u$, we
generate a curved space null cone with the same conformal structure,
i.e.  $\gamma_{ab}=\Omega^2\hat \gamma_{ab}$. We thus set $R=\Omega
\hat R$ and $h_{ab}={\hat h}_{ab}$. We do not require that the two
affine structures agree and set $n^a=\Lambda \hat n^a$ so that
$\partial_u = \Lambda \partial_{\hat u}$ and $\Sigma =\Lambda \hat
\Sigma$. Our goal is to investigate the properties of the foliation
$S_t$ determined by translating ${\cal S}_0$ through the curved space
affine time $\Delta u =t$.

The curved space focusing equation (\ref{eq:focus}) now
reduces to
\begin{equation}
    \Lambda '(\Omega ' \hat R+\Omega \hat R')
    +\Lambda(\Omega ''\hat R +2\Omega '\hat R')=0;
\label{eq:foc1}
\end{equation}
and the Weyl curvature, defined in Eq. (\ref{eq:psiab}), reduces to
\begin{equation}
       \Psi_{ab} = \frac{\sigma \Lambda}{2}(\Omega^2\Lambda)'
		   (p_a p_b -q_a q_b)
\end{equation}
with the Weyl scalar, defined in Eq. (\ref{eq:Psi}), given by
\begin{equation}
     \Psi = \sigma \frac{ \Lambda (\Omega^2\Lambda)'}{\Omega^2 \hat R^2}.
\label{eq:psi}
\end{equation}

The goal is to solve these equations to construct a non-singular white
hole horizon ${\cal H}$. Then $\Psi$ must be non-singular on ${\cal
H}$ and the scalar fields $\Omega$ and $\Lambda$ must be smooth
positive functions, except possibly at the endpoints of ${\cal H}$. In
addition, the surface area of ${\cal H}$  must approach a finite limit
as $u \rightarrow -\infty$.

We require that $u$ and $\hat u$ approach $-\infty$ together at the
same rate so that we may restrict the affine freedom in $u$ by
requiring that $u' \rightarrow 1$.  The surface area function must have
a finite limit $R \rightarrow R_{\infty}$ as $u \rightarrow -\infty$,
corresponding to an irreducible mass~\cite{wald}
$M_{\infty}=R_{\infty}/2$. Then, from inspection of Eq. (\ref{eq:psi}),
$\Omega \hat u \rightarrow -R_{\infty}$ and $\Lambda \rightarrow 1$, as
$\hat u \rightarrow -\infty$. We also assume that these conditions are
uniformly satisfied along each null ray in terms of a $1/\hat u$
expansion. This puts constraints on the fields $\Lambda$ and $\Omega$
which satisfy Eq.  (\ref{eq:foc1}). In order to apply these conditions
it is convenient to introduce the function $F=\Lambda\Omega^2$. Then
the smoothness of $\Psi$ requires that $F'=0$ at a caustic. Also, the
asymptotic conditions on $\Omega$ and $\Lambda$ as $\hat u \rightarrow
-\infty$ require $\hat u^2 F\rightarrow R_{\infty}^2$ so that, $\hat u
(log F)' \rightarrow -2$.

Equation (\ref{eq:foc1}) can now  be rewritten in terms of $F$ and
$\Omega$ as
\begin{equation}
      (log F)'=\frac{\Omega[1/\Omega]''}{[log(\Omega \hat R)]'}.
\label{eq:foc2}
\end{equation}
We can generate solutions to Eq. (\ref{eq:foc2}) by making an ansatz
for $\Omega$ and then integrating to determine $F$. The above
smoothness condition that $F$ must satisfy at a caustic is then
automatically satisfied if $\Omega$ is smooth. The asymptotic
conditions require that the ansatz satisfy $\Omega \hat R \rightarrow
R_{\infty}+O(1/\hat u^2)$ and $u^2(\Omega \hat R
-1)\rightarrow -\sigma^2 /24$.

In order for the resulting model to represent a non-singular white
hole, additional conditions arise at a shear-free ray. Along such a
ray, the focusing equation (\ref{eq:focus}) implies that $\partial_u^2
R =0$ with solution $R=C_1+C_2 u$. Accordingly, $R$ must be a constant
along each ray since it approaches a finite limit $R_{\infty}$ as $u
\rightarrow -\infty$. Thus we must require that our ansatz reduce to
$\Omega =R_{\infty}/\hat R$ along a shear-free ray.  Such rays occur at
any umbilical point of ${\cal S}_0$ where the two curvature eigenvalues
are equal and $\sigma =0$. For surfaces of revolution the poles are
always umbilical. Umbilics are a major factor in determining the
qualitative behavior of the white hole model. Along a non-umbilical
ray, the completeness of ${\cal H}$ as a white hole model requires that
the range of $u$ extend to a crossover point or caustic, where it hits
another ray and the hole terminates. However, along an umbilic, the
white hole need not terminate and can extend to infinite $u$. As we
shall illustrate, this is the mechanism which leads to multiple black
holes in a $u$-foliation of ${\cal H}$.

The behavior at umbilics imposes further conditions on our ansatz. For
example, suppose that the white hole terminates along a set of
non-umbilical caustics $\hat u=-\sigma/2$ which has an endless umbilic
ray on its boundary where $\sigma =0$. Then $u$ must be finite along
the non-umbilic set but approach $\infty$ as $\sigma \rightarrow 0$.
Also, $u'= 1/\Lambda$ must have the same behavior since the caustic
set, including its boundary, is reached at finite $\hat u$.

The simple ansatz
\begin{equation}
    \Omega=-R_{\infty}\big( \hat u
	  +\frac{\sigma^2}{12(\rho-\hat u )}\big)^{-1}
\label{eq:ansatz}
\end{equation}
satisfies all the above conditions if the parameter $\rho$ is chosen so
that $\rho \ge \sigma/\sqrt{13}$. Then $\Omega >0$ in the white hole
region contained inside $\hat u \le -\sigma /2$. Furthermore,
integration of Eq. (\ref{eq:foc2}) gives
\begin{equation}
     F=\frac{16 R_{\infty}^2 (\hat u-\rho)^2 \,
	 (2 \hat u -5 \rho + \mu)^{2 \,
      (2\rho/\mu -1) \,} }{(2 \hat u -5 \rho - \mu)^{2 \,
    (2\rho/\mu +1) \,}}
\end{equation}
and
\begin{equation}
    u'= \frac{9}{(12 \hat u (\hat u-\rho) -
      \sigma^2)^2} \frac{(2 \hat u -5 \rho - \mu)^{2 \,
	  (2\rho/\mu +1) \,}}{(2 \hat u -5 \rho+ \mu)^{2 \,
	  (2\rho/\mu -1) \,} } \, ,
\label{eq:lampr}
\end{equation}
where $\mu = \sqrt{13\rho^2 -\sigma^2}$ is real and positive.
The asymptotic expansion of the integral gives
\begin{equation}
u = {\it \hat u}-12\,\rho\,\ln {\it \hat u} +C+ O(\frac{1}{\hat u}) ,
\label{eq:lamasym}
\end{equation}
where $C$ is the integration constant.

On the caustic set $\hat u=-\sigma /2$, we have
\begin{equation}
 \Psi=\frac{8 \sigma^4}{81 (\sigma + 2 \rho)^6 (\sigma + 3 \rho)}
 \Bigg( \frac{ \sigma +5 \rho - \mu }{ \sigma + 5 \rho
 +\mu} \Bigg) ^{8\rho/\mu}  \; ,
\end{equation}
which is manifestly regular;
and
\begin{equation}
 u'=\frac{9 (\sigma + 2 \rho)^2}{\sigma^2} \Bigg( \frac{  \sigma
 +5 \rho + \mu}{ \sigma +5 \rho - \mu} \Bigg )^{4\rho/\mu} \, ,
\end{equation}
which displays the required singular behavior as $\sigma \rightarrow 0$
at an umbilical point and is otherwise regular.

It is instructive to examine the behavior of the extrinsic
curvature eigenvalues defined on ${\cal H}$ according to
\begin{equation}
     \kappa_p =\frac {p^ap^b{\cal L}_n \gamma_{ab}}{2R^2}
		=\frac {\Sigma}{2} +\frac{\partial_u R} {R}
\end{equation}
and
\begin{equation}
     \kappa_q =\frac{ q^aq^b{\cal L}_n \gamma_{ab}}{2R^2}
		=-\frac {\Sigma}{2}+\frac{\partial_u R}{R}.
\end{equation}
For our flat space model, $\hat \kappa_p =1/(\hat u-\sigma /2)$ and
$\hat \kappa_q =1/(\hat u+\sigma /2)$. Then $\kappa_{(p,q)} =
\Lambda(\hat \kappa_{(p,q)}+\Omega^{-1} \partial_{\hat u} \Omega)$.
For the ansatz (\ref{eq:ansatz}), these reduce to
\begin{eqnarray}
     \kappa_p &=& \frac{\sigma (12 u (u -\rho) - \sigma^2)
	   ( 12 (u -\rho)^2 + \sigma (\sigma- 4 u + 2 \rho) )}
	   {9 (u -\rho) (2 u-\sigma)}
	   \frac{(-2u+5\rho-\mu)^{2[(2\rho/\mu)-1]}}
	  {(-2u+5\rho+\mu)^{2[(2\rho/\mu) +1]}}
\label{eq:kappap}
\end{eqnarray}
and
\begin{eqnarray}
     \kappa_q &=& \frac{-\sigma (12 u (u -\rho) - \sigma^2)
	    ( 12 (u -\rho)^2 + \sigma(\sigma+ 4 u - 2 \rho) )}
       {9 (u -\rho) (2 u + \sigma)}
   \frac{(-2u+5\rho-\mu)^{2[(2\rho/\mu)-1]}}
      {(-2u+5\rho+\mu)^{2[(2\rho/\mu)+1]}}.
\label{eq:kappaq}
\end{eqnarray}
Both $\kappa_p$ and $\kappa_q$ approach 0 as the white hole approaches
``equilibrium'' as $u \rightarrow -\infty$. However, for large
negative $u$ it is clear from examining the dominant terms in
Eq's (\ref{eq:kappap}) and (\ref{eq:kappaq}) that $\kappa_p >0$ and
$\kappa_q <0$. So, although the mean curvature of ${\cal H}$ is
negative in agreement with net focusing, ${\cal H}$ expands in
the $p$ principle direction. As illustrated in Sec. \ref{sec:axi}, the
effect of this expansion as it becomes stronger near the $q$-caustic
provides the characteristic shape of a bifurcating horizon.

\section{The conformally spheroidal case}
\label{sec:axi}

The preceding analysis describes the dependence of the null geometry of
${\cal H}$ on affine parameter along a ray. In order to obtain a model
of an exploding white hole we examine the global dependence of the
geometry on the angular coordinates parameterizing the rays of ${\cal
H}$. In the simplest case, ${\cal S}_0$ is a sphere. Then $\sigma =0$
on all rays, $\hat h_{ab}=q_{ab}$ the unit sphere metric, $\hat R
=-\hat u$ and the ansatz Eq.(\ref{eq:ansatz}) reduces to $R=R_{\infty}$
so that the horizon is stationary. A conformally spherical white
(black) hole is topless (bottomless) along all rays.

Since all topologically spherical geometries are conformally related,
the conformal null geometry of a stationary white (or black) hole is
unique. What distinguishes a  Kerr horizon from a Schwarzschild horizon
is the initial geometry of ${\cal S}_0$, which determines $R_{\infty}$.
In the Kerr case, $R_{\infty}$ is the conformal factor relating the
geometry of the Kerr horizon to the unit sphere. (The spin of the Kerr
hole arises from an extrinsic geometric quantity which can also be
freely specified on ${\cal S}_0$). In the Schwarzschild case,
$R_{\infty}$ can chosen to be a constant (independent of angle).

In the case in which ${\cal S}_0$ is a prolate spheroid (ellipsoid of
revolution), we will show how the identical horizon structure arises,
as found in the simulation of colliding black holes. Furthermore, in
the oblate case,  we will show how a temporarily toroidal horizon
arises, as found in the simulation of rotating collapse.

We start with the $2$-dimensional surface ${\cal S}_0$ describing the
spheroid
\begin{equation}
\frac{x^2+y^2}{a^2}+\frac{z^2}{b^2}=1 \, ,
\end{equation}
which can be alternatively described in coordinates
$y^A = (\theta, \phi)$, by the map
\begin{eqnarray}
x&=&a \sin\theta \cos\phi \, \\
y&=&a \sin\theta \sin\phi \, \\
z&=&b \cos\theta \, ,
\end{eqnarray}
with $ \theta \in [0, \pi]$ and $\phi \in [0, 2 \pi )$.
The intrinsic metric of ${\cal S}_0$ is
\begin{equation}
{\hat g}_{AB} dy^A dy^B =(a^2 \cos^2 \theta + b^2 \sin^2 \theta )
   d\theta^2 +  a^2 \sin^2 \theta d\phi^2 \, .
\end{equation}
The determinant condition provides a way to define ${\hat R}^2$ as
\begin{equation}
{\hat R}^2 = \frac{ det({\hat g})}{det(q)},
\end{equation}
(with $q_{AB}$ the unit sphere metric in $(\theta,\phi)$ coordinates).
Therefore, ${\hat R}^2 = a \sqrt{b^2 \sin^2 \theta +a^2 \cos^2 \theta
}$. A straightforward calculation provides
the principal radii of curvature of ${\cal S}_0$,
\begin{equation}
r_{\theta}=\frac{(a^2 \cos^2 \theta +b^2 \sin^2 \theta )^{3/2}}{ab}
\end{equation}
and
\begin{equation}
r_{\phi} = \frac{a \sqrt{a^2 \cos^2 \theta +b^2 \sin^2 \theta}}{b} \, .
\end{equation}

We consider ${\cal S}_0$ to be isometrically embedded at time
$t=0$ in Minkowski space and identify the $\hat u$ foliation of
${\cal H}$ with the (ingoing) null hypersurface emanating from ${\cal
S}_0$.  We invariantly identify the function $\sigma$ in our model
of ${\cal H}$ as the difference between the principle curvatures of
${\cal S}_0$:
\begin{equation}
\sigma = |r_{\theta} - r_{\phi}| =
       \frac{|b^2-a^2| \sin^2 \theta
	 \sqrt{a^2 \cos^2 \theta +b^2 \sin^2 \theta }}{a b}.
\end{equation}
We set $\hat u = u_0$ on ${\cal S}_0$. Then our
convention that $\hat u =0$ midway between the two caustics of
${\cal H}$ allows us to invariantly identify
\begin{equation}
  u_0 =-\frac {(r_{\theta} + r_{\phi})}{2} =
    -\frac{(2a^2+(b^2-a^2)\sin^2 \theta)
      \sqrt{a^2 \cos^2 \theta +b^2 \sin^2 \theta }}{2a b}.
\label{eq:u0}
\end{equation}
This determines both the metric $\hat \gamma_{ab}$ and affine parameter
$\hat u$ for the flat spheroidal null cone.

The conformal factor for the metric $\gamma_{ab}=\Omega^2 \hat
\gamma_{ab}$ of the curved space version is determined by the ansatz
Eq.(\ref{eq:ansatz}). Here, in order to satisfy smoothness conditions,
we require that the parameter $\rho \ge \sigma_M/\sqrt{13}$, where
$\sigma_M$ is the maximum value of $\sigma$ attained on
${\cal S}_0$. For $a \le b \sqrt{3}$, the maximum occurs at the
equator and $\sigma_M = |b^2-a^2|/a$. For the highly oblate case with
$a > b \sqrt{3}$, the maximum occurs
between the equator and poles and $\sigma_M = 2a^2/(3\sqrt{3}b)$.

To determine the curved space affine
parameter, we set $u=u_0$ on ${\cal S}_0$. Along with the condition
that $\partial u/ \partial \hat u \rightarrow 1$ as $\hat u \rightarrow
-\infty$ this fixes the remaining affine freedom in $u$, i.e. the ray
dependent integration constant $C$ in Eq.  (\ref{eq:lamasym}). With
this choice, $u =\hat u$ on ${\cal S}_0$ and $\partial_u
/\partial_{\hat u}\rightarrow 1$ as $\hat u \rightarrow -\infty $.

Figure~\ref{fig:null} illustrates the features of the flat space spheroidal
null
hypersurface ${\cal N}_{flat}$, with horizontal lines corresponding to the
foliation $\hat S_t$ given by $\hat u = u_0 +t$. In the illustration,
we suppress the rotational symmetry. As discussed below, when the
spheroid is prolate (oblate), the crossover points $X$ where ${\cal
N}_{flat}$ pinches off is a space-like line (disc).  The features of the
conformally related curved space version ${\cal N}$ are quite similar
when viewed with respect to the Minkowski foliation $\hat S_t$. The
chief difference is the effect of the conformal factor on the expansion
and shape of the surfaces $\hat S_t$, which produces a finite surface
area as $\hat u \rightarrow -\infty$. The features of ${\cal N}$ with
respect to the curved space affine foliation $S_t$, given by $u = u_0
+t$, are qualitatively similar to those for $\hat S_t$ at early times.
The interesting black hole physics occurs near the crossover region of
where the foliations $S_t$ and $\hat S_t$ have topologically different
properties. These are best illustrated by embedding techniques.

\subsection{Embedding}

Embedding diagrams constitute valuable tools for visualizing the
intrinsic geometry of a curved 2-dimensional surface. By embedding the
surface in a flat 3-dimensional Euclidean space, one obtains a surface
with the same intrinsic geometry. The following technique was developed
by Smarr, who applied it to the description of the Kerr black
hole~\cite{smarr_emb}.  More recently, it has been employed to analyze
the event horizon of the head-on-collision of black
holes~\cite{masso}.  Here we describe its application to our model.

The first step is to introduce the angular coordinate $\eta=\cos
\theta$ which makes $det(q_{AB})=1$.  In $(\hat u,\eta,\phi)$
coordinates, the intrinsic metric of the horizon is
\begin{equation}
\gamma_{ab}dx^a dx^b = \Omega^2 \hat R^2 (  f^{-1} d\eta^2 + f d \phi^2)
\end{equation}
where
\begin{equation}
     f = \frac{(1-\eta^2) (\hat u + \sigma/2)}{(\hat u -\sigma/2)}.
\end{equation}
This can then be transformed into $(u,\eta,\phi)$ coordinates by the
substitution $\hat u = \hat u(u,\eta,\phi)$, where $\hat
u(u,\eta,\phi)$ is determined by integrating Eq. (\ref{eq:lampr}). The
results of this paper are based upon carrying out the integral by means
of a Taylor expansion in $u$ about $u_0$ up to 6th order. (No
substantial change in our results were seen in going from 4th order to
6th).

Now, one can isometrically embed this surface in a 3-dimensional
Euclidean space with Cartesian coordinates $x^i$ by the map
\begin{equation}
x^1 = F(\eta) \cos\phi, \, x^2 = F(\eta) \sin\phi, \, x^3 = G(\eta) ,
\end{equation}
where
\begin{eqnarray}
F &=& \Omega \hat R \sqrt{f} \nonumber \\
G_{,\eta} &=& \sqrt{ \Omega^2 \hat R^2/f - F_{,\eta}^2 } .
\end{eqnarray}

The quantities $F$ and $G$ are used to display the surface in the
familiar 3-dimensional flat space at a given instant of time determined
by the $u$ foliation. Moreover, one can monitor the embedding at
different instants of time and produce an ``embedding history'' which
shows the evolution of the surface's geometry. By suppressing the $\phi$
direction one can stack $\phi=const$ cross-sections of the embedding
of the $S_t$ foliation in a three-dimensional fashion, with the
vertical axis labeling $t$ and the horizontal axes labeling
$F=\sqrt{(x^1)^2+(x^2)^2}$ and $G=x^3$.

\subsection{The pair-of-pants}

\begin{figure}
\centerline{\epsfxsize=6in\epsfbox{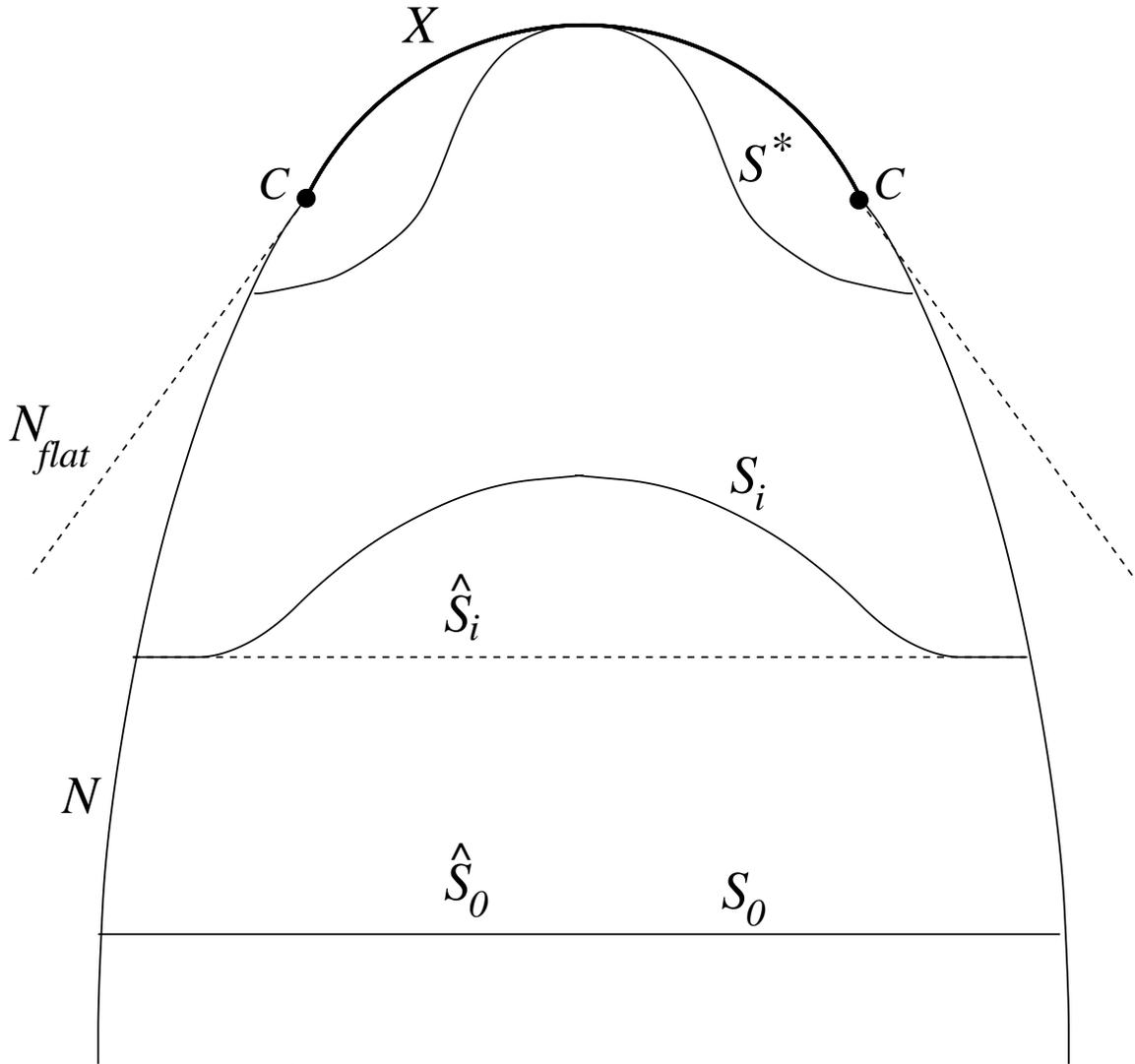}}
\caption{Spheroidal (${\cal N}_{flat}$) and conformally spheroidal
(${\cal N}$) null hypersurfaces: Factoring out the rotational
symmetry allows the foliations to be depicted as lines from pole to
pole. The Minkowski foliation indicated by $\hat S_i$ is drawn
horizontally. The curved affine foliation is indicated by $S_i$.}
\label{fig:null}
\end{figure}

We first describe the prolate case $b>a$ in which the crossover points
$X$ in the flat-space model are also a line of caustics with respect to
the $\phi$ principle direction. (Thus the $\phi$ direction corresponds
to the $q$ principle direction). In Fig.~\ref{fig:null} the rotational
symmetry has been factored out so a Minkowski time foliation $\hat S_t$
of the underlying flat space spheroidal null hypersurface corresponds
to horizontal lines. The effect of curvature focuses the conformally
related null hypersurface ${\cal N}$ and introduces an upward bulge in
the $S_t$ foliation, which gets enhanced at later times to produce the
slice $S^*$ at which the white hole bifurcates. The vertical time
sequence in our figures corresponds to white holes but the figures
can be turned upside-down to depict the corresponding scenario for
black holes, in this case a black hole merger. The points $C$ represent
the caustics at the poles which are reached at finite times in the flat
model but correspond to infinite affine times in the curved model.

\begin{figure}
\centerline{\epsfxsize=6in\epsfysize=5.7in\epsfbox{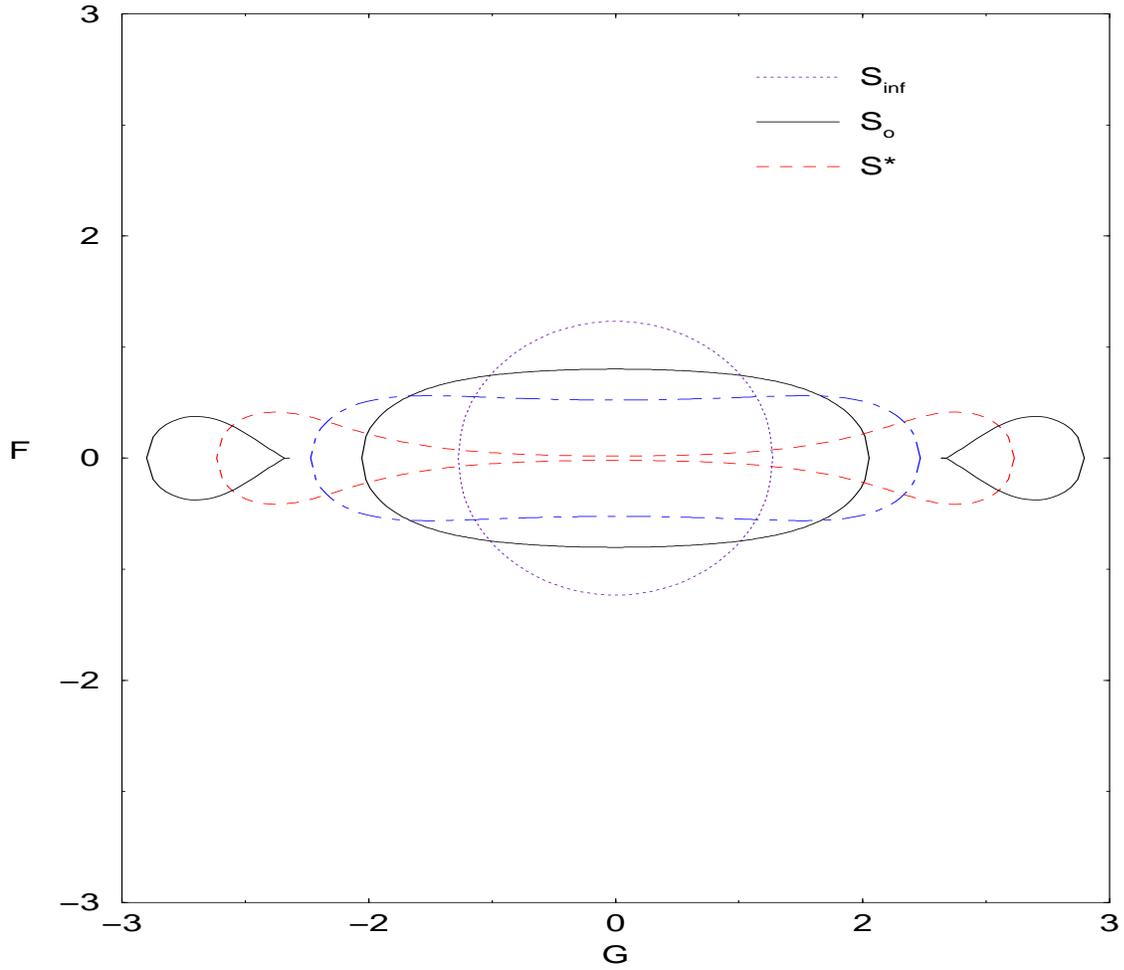}}
\caption{Embedding snapshots profiling fission of an initially
spheroidal white hole ${\cal S}_0$ in terms of the coordinates $F$ and
$G$: The white holes are the surfaces formed by rotating the profiles
about the horizontal $G$ axis.}
\label{fig:snap}
\end{figure}

Profiles of the embedding diagram of $S_t$ at various stages are
shown in Fig.~\ref{fig:snap}. Proceeding backward in time from the
initial prolate spheroid ${\cal S}_0$, the cross-sections form the
sphere $S_{inf}$ as $t\rightarrow -\infty$. Proceeding forward in
time, they form the surface $S^*$ (also indicated in Fig.~\ref{fig:null})
where the white hole is at the verge of fissioning. Note that the two
white holes which are produced each have a sharp point at their inner
pole.

\begin{figure}
\centerline{\epsfysize=6in\epsfbox{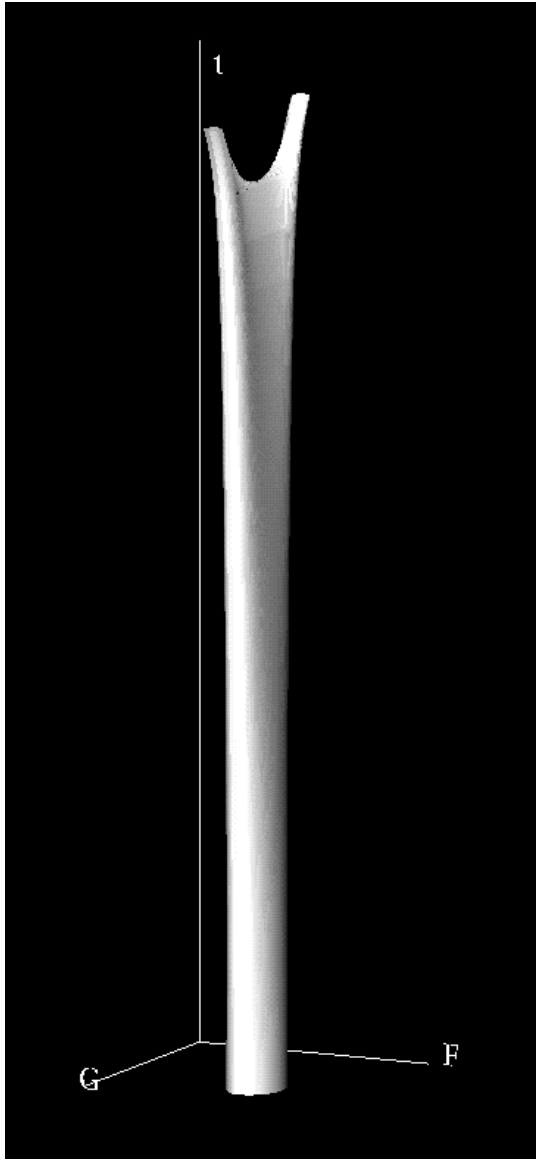}}
\caption{The embedding history of the axisymmetric fission of a white
hole into two white holes: The time slices $S_t$ are horizontal and
proceed upward as $t$ increases..  The spatial origin of the embedding
axes is offset from the center for clarity. The suppressed symmetry
dimension corresponds to a rotation about the $G$ axis. The history
extends into both the future and past of the initial surface ${\cal
S}_0$, which lies approximately halfway up the picture. A time reversed
view gives the pair-of-pants picture for the head on collision of two
black holes.}
\label{fig:pants}
\end{figure}

\begin{figure}
\centerline{\epsfxsize=6in\epsfbox{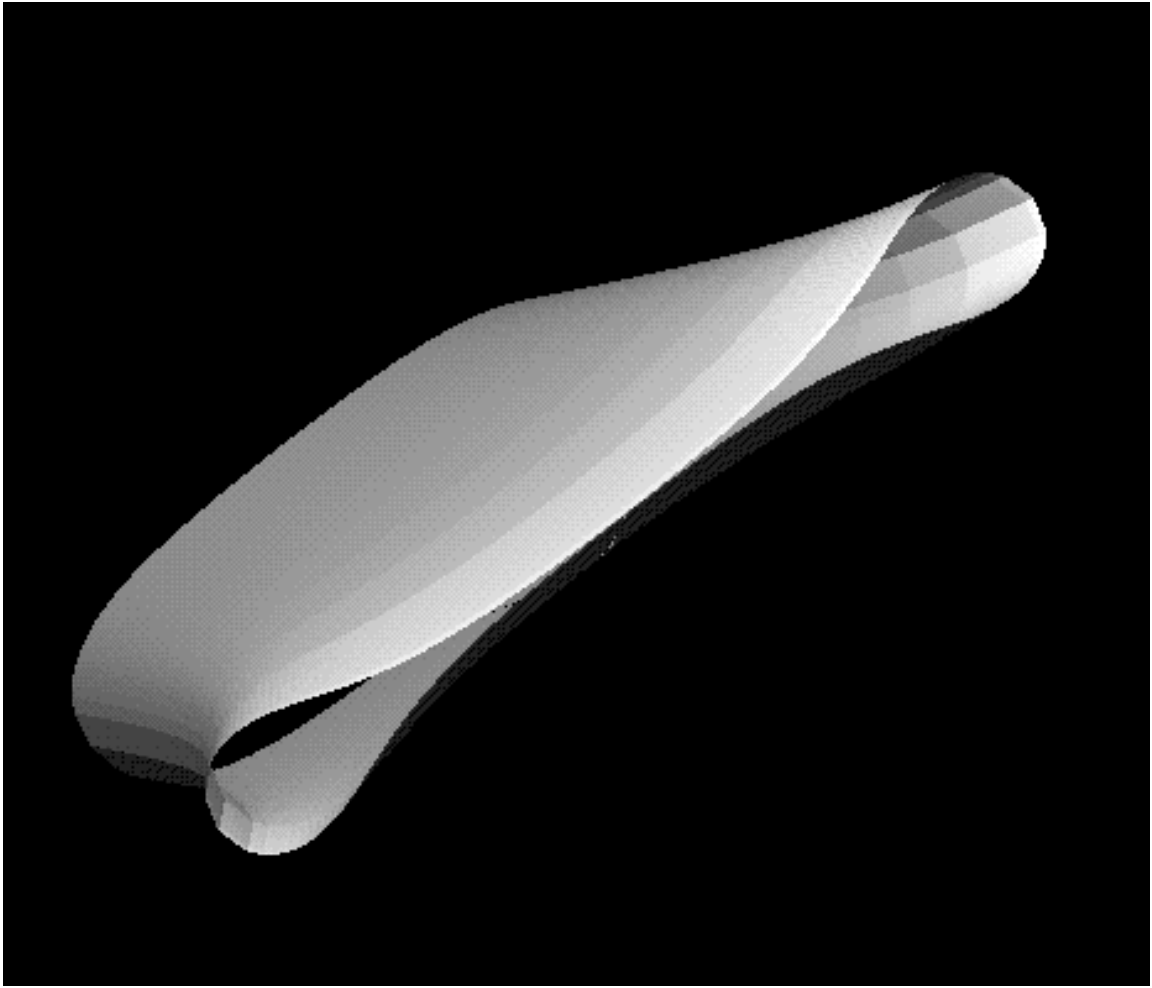}}
\caption{A downward view of the portion of the embedding history
from ${\cal S}_0$ up to the verge of bifurcation where the light rays
at opposite points on the equator are about to cross.}
\label{fig:crotch}
\end{figure}

Fig.~\ref{fig:pants} shows a time stacking of embedding diagrams of the
$S_t$ foliation to form an (inverted) pair-of-pants and
Fig.~\ref{fig:crotch} gives a cutaway view  of the bifurcation.  The
main features of the pair-of pants agree with those found in numerical
simulation of the head-on vacuum black hole collision~\cite{annin1}, as
described in Ref's~\cite{sci,masso}.  However, the analytic nature of
the present work allows us to draw the following further conclusions.
First, for the vacuum case, the pair-of-pants is eternal along the two
umbilical rays at the poles. However, the legs pinch off and shrink
asymptotically since every other ray eventually reaches the crossover
$X$ at finite $u$. Also, referring to the discussion following Eq.
(\ref{eq:kappaq}), the principle curvature $\kappa_q$ corresponding to
the $\theta$ direction is everywhere negative, except at the poles and
in the limit $\hat u \rightarrow -\infty$ where $\kappa_q=0$. It is
most negative along the equatorial rays, which gives rise to the
bow-legged shape of the pair-of-pants.

\subsection{Toroidal horizons and the hoop conjecture}

In the oblate case $a>b$, the crossover points $X$ in Fig.~\ref{fig:null}
(in which the orbits of the rotational symmetry have been factored out)
correspond to the same crossover points as in the prolate case. The
difference between the two cases is the orientation of the axis of
rotation. Whereas $X$ lies on the rotation axis in the prolate case
(and thus determines a caustic line under revolution), in the oblate
case $X$ rotates to form a disc. Only the outer rim of the disc
(generated by revolution of the equatorial points $C$ in
Fig.~\ref{fig:null}) consists of caustic points.

Note that the induced metric $\gamma_{ab}|_X$ of the crossover disc is
single-valued (except at the caustic rim $C$ where it is singular);
i.e, its value does not depend upon whether $X$ is approached from the
top or bottom. This is because (i) by construction of ${\cal N}$ as a
null hypersurface embedded in Minkowski space, $\hat \gamma_{ab}|_X$ is
single-valued and (ii) the conformal factor $\Omega$ has reflection
symmetry with respect to the equatorial plane. More generally, in the
absence of symmetry, establishment of consistency conditions for a
single-valued metric on the crossover surface would be more
complicated.

In the oblate case the $\theta$ direction has the smallest radius of curvature
(corresponding to the $q$ principle direction). The umbilical rays at the poles
cross before they caustic, so that  the infinite umbilical stretch in the $S_t$
foliation for the prolate case does not arise in the oblate case. Thus the
event horizon is completed in finite affine time, in contrast to the prolate
case.

Another important difference also arises in this case. The umbilical
stretch produces toroidal cross-sections of the horizon, rather than
the two spherical sections arising in the prolate case. The details
of the formation of the
torus are best understood in terms of the affine displacement $\Delta
u(\theta)$ between ${\cal S}_0$ and $X$, as a function of the
$\theta$-coordinate of the ray. In terms of Minkowski time, the
corresponding time displacement is~\cite{toroid}
\begin{equation}
      \Delta \hat u(\theta) = b^2 \sqrt{\frac {\sin^2\theta}{a^2}
		  +\frac {\cos^2\theta}{b^2} } \, .
\end{equation}
Then
\begin{equation}
      \Delta u(\theta) = \int_{u_0}^{u_0+\Delta \hat u} u' d\hat u\, ,
\end{equation}
where $u'$ is given by Eq. (\ref{eq:lampr}) and $u_0$ by Eq.
(\ref{eq:u0}). In the oblate case, $\Delta \hat u$ has its minima at the
equator but $\Delta u$ has minima at the poles and exhibits
a monotonic growth towards the equator. As a result,
the $S_t$ foliation first touches $X$ at the space-time point where the
two polar rays intersect, creating a surface of revolution with
the same double teardrop profile as $S^*$ in Fig.~\ref{fig:snap}, but
now rotated about the vertical axis through the center. In successive
cross-sections, $S_t$ forms a torus (with sharp inner rim), which
shrinks to a circle as the horizon terminates.  The tidal deformation
introduced in the $S_t$ foliation of the curved space model is somewhat
analogous to the pair-of-pants shape, except now the identifications of
the suppressed rotational symmetry lead to a toroidal topology of $S_t$
for a period of time following the bifurcation. This regime corresponds
to the scenario found in the numerical simulation of the collapse of a
rotating cluster of particles~\cite{shteuk,torus,toroid}.

\begin{figure}
\centerline{\epsfxsize=6in\epsfbox{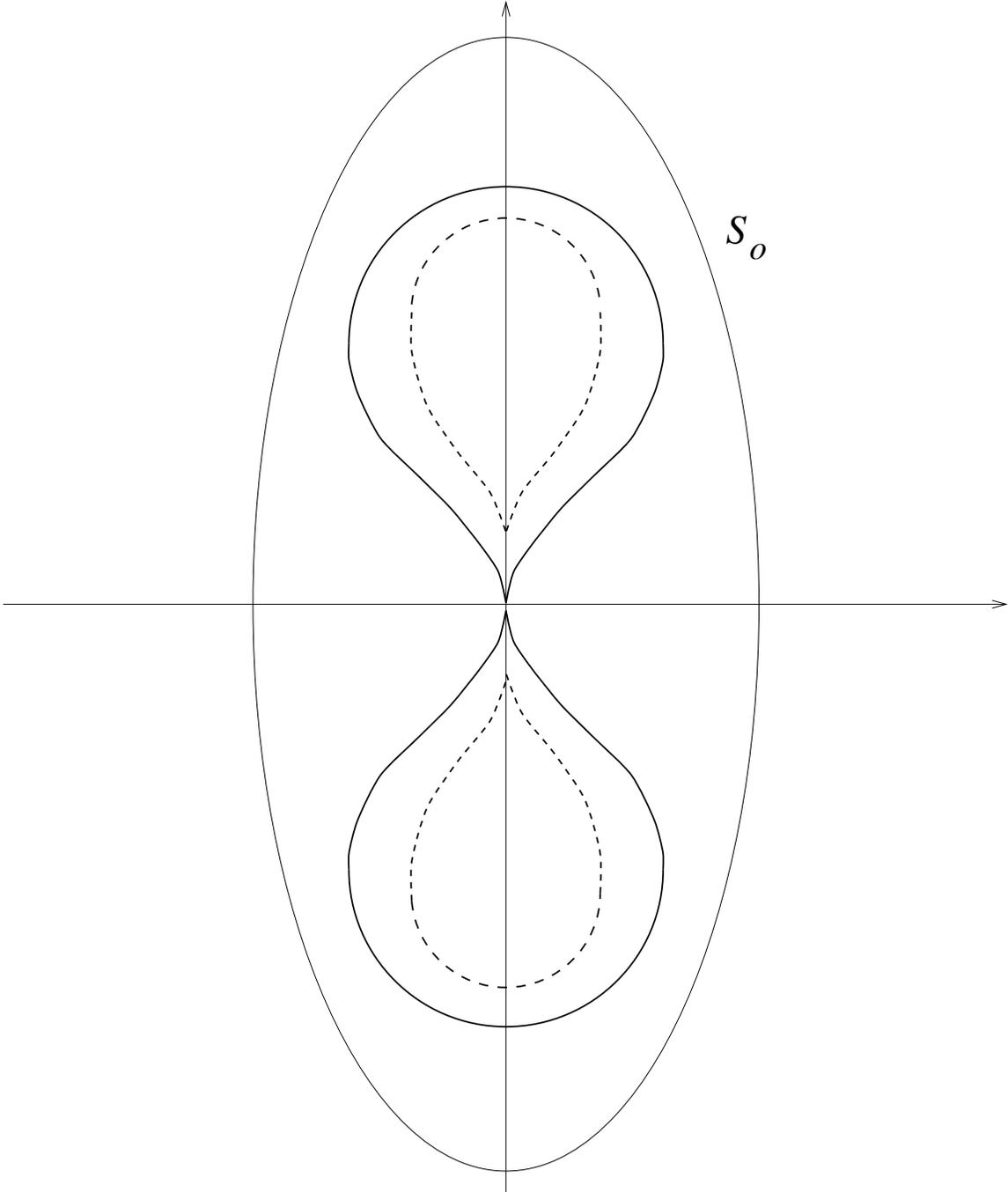}}
\caption{Schematic profiles of an initially spheroidal white hole $S_0$
going through a toroidal stage. The rotation axis is vertical. The heavy
solid line depicts the shape just prior to formation of the torus. The
inner rim of the torus  (dashed line) has a non smooth edge.}
\label{fig:tor}
\end{figure}

Euclidean embedding is not possible for this full sequence of toroidal white
hole formation. Similar results were previously noted by Smarr~\cite{smarr_emb}
in regard to the non-existence of a Euclidean embedding for high angular
momentum Kerr black holes. The schematic profiles in Fig.~\ref{fig:tor}
indicate the qualitative topological features of the evolution.

The analytic nature of the present approach allows us to draw further
conclusions. In the oblate case, the $p$ principle curvature direction
in which $\kappa_p$ is positive corresponds to the $\phi$ direction, so
that the equatorial circumference $C$ is always larger than its
asymptotic value $2\pi R_{\infty}$. This has important bearing
on  the hoop conjecture~\cite{hoop}, which in its original formulation
would require that $C\lesssim 4\pi M$. The exact nature of the mass
$M$ and of the inequality were purposely left vague in the statement of
the conjecture for purposes of further mathematical refinement. If we
identify $M$ as the irreducible mass associated with the surface area
$4\pi R_{\infty}^2$ then $M=R_{\infty}/2$ and $C> 4\pi M$ at every
finite $u$. The largest value of $C$ occurs at the equatorial rim of
the crossover disc where
\begin{equation}
     C_X =\frac {3(\sqrt{13}+2)}{(\sqrt{13}+3)}[4\pi M]
	   \approx 2.546 [4\pi M] ,
\label{eq:cx}
\end{equation}
for the choice $\rho =\sigma_M$ (which maximizes the result).  Although
our model should not be expected to provide the sharpest bound, this
result suggests a significant restriction on any plausible version of the
hoop conjecture.

\section{Conclusion}

We have shown that it is possible to treat multiple black or white
holes via a stand-alone-model of the event horizon based upon
constraint equations for the characteristic initial value problem. In
this paper, we have concentrated on the constraint governing the
internal geometry of the horizon. Remarkably, this single equation
produces such rich results. Even more interesting features should be
expected for models conformal to flat space null hypersurfaces with
more structure than the spheroidal case considered here.

In subsequent work, we will extend the treatment to the constraint
governing the extrinsic curvature. The boundary conditions provided by
the solution of this constraint problem is the missing ingredient
necessary to evolve the exterior space-time by means of an existing
characteristic code.

\acknowledgements

This work has been supported by NSF PHY 9510895 and NSF INT 9515257 to the
University of Pittsburgh and by the Binary Black Hole Grand Challenge Alliance,
NSF PHY/ASC 9318152. N.T.B. thanks the Foundation for Research Development,
South Africa, for financial support, and the University of Pittsburgh for
hospitality. L.L. thanks the Universities of South Africa and of
Durban-Westville for their hospitality. Computer time for the graphical
representations was provided by the Pittsburgh Supercomputing Center under
grant PHY860023P. We thank Joel Welling of the PSC for assistance with the
visualizations. We are especially grateful to Sascha Husa for important
discussions and a critical reading of the manuscript.

\newpage

\newpage

\newpage

\newpage

\newpage

\newpage

\end{document}